\documentclass[12pt]{article}
\usepackage{amsfonts}
\usepackage{fancybox}
\usepackage[dvips]{graphicx}
\makeatletter
\@addtoreset{equation}{section}

\makeatother
\begin{document}
\begin{titlepage}
 
$\mbox{ }$
\begin{flushright}
\begin{tabular}{l}
\\
arXiv:0803.3167 [physics.ed-ph]\\
TIFR/TH/07-46\\
IU-TH-5
\end{tabular}
\end{flushright}
~~\\
~~\\
~~\\

\vspace*{0cm}
    \begin{Large}
       \vspace{2cm}
       \begin{center}
        An inquiry into the reproduction of physics-phobic children \\
        by physics-phobic teachers\footnote{English translation of the original 
        version written in Japanese \cite{kiyou}}
       \end{center}
    \end{Large}

  \vspace{1cm}

\begin{center}
          Takehiro A{\sc zuma}$^{\dag}$ \footnote{
e-mail address : azuma@theory.tifr.res.in} \footnote{Address after Apr. 1, 2008: Department of Mathematics and Physics, Setsunan University, 17-8 Ikeda Nakamachi, Neyagawa, Osaka, 572-8508, Japan, e-mail address: azuma@mpg.setsunan.ac.jp } and
          Keiichi  N{\sc agao}$^{\ddag}$ \footnote{
e-mail address : nagao@mx.ibaraki.ac.jp }
 \\

\vspace*{1cm}
$^{\dag}$ {\it Department of Theoretical Physics, Tata Institute of Fundamental Research, Homi Bhabha Road, Mumbai, 400005, India} \\
\vspace*{1cm} 
$^{\ddag}$ {\it Theoretical Physics Laboratory, Faculty of Education, Ibaraki University, Mito 310-8512, Japan}
\end{center}

\vfill

\begin{abstract}
\noindent
Recently in the authors' country Japan, the unpopularity of natural science 
among children has been a serious problem. 
Especially, physics is unpopular because physics requires mathematics. 
One of the reasons of this problem is that teachers themselves do not like physics. 
We focus our attention on the ``teachers in embryo'', 
namely the undergraduate students in 
a course for school teachers. 
We conducted a questionnaire and a quiz on the undergraduate students 
in the first grade of the Department of Science Education, Ibaraki University. 
We report the result of the questionnaire and the quiz, 
and also make suggestions to improve the present situation.

\end{abstract}
\vfill
\end{titlepage}
\vfil\eject

\tableofcontents

\section{Introduction}
Recently in the authors' country Japan, the unpopularity of natural 
science among children has been a serious problem. 
One of the reasons of this phenomenon is the unpopularity of natural science 
among the teachers. Since physics makes use of a lot of advanced mathematics, 
physics is particularly unpopular. 
According to the report of the Japanese Ministry of Economy, 
Trade and Industry \cite{keisansho}, 
60 $\%$ of the students aiming for 
teaching natural science did not study physics 
when they were high school students, and less than 20 $\%$ of them like physics.

The unpopularity of physics is not only the matter of incompetence in natural science 
but also a serious problem that affects the future of the nation in the long run. 
We are now urged to figure out the reason and improve the present situation. 
We thus conducted a survey about the academic ability 
of and the interest in physics, 
on the undergraduate students in the first grade of the Department of Science 
Education, Ibaraki University in July 2006. 
%
%
We propose a plan to solve the problem of physics unpopularity. 
This paper is organized as follows. 
In Section \ref{chap-ques}, we present 
the result of the survey. 
In Section \ref{chap-plan}, we make
suggestions to improve the present situation. 
Section \ref{chap-conclu} is devoted to the summary. 
We show the Japanese teaching guidelines 
of high school mathematics and physics in Appendix \ref{app-math}, and  
the detail of the quiz and the questionnaire in Appendix \ref{app-test}.

\section{Quiz and questionnaire about physics} \label{chap-ques}
In this section, we report the result of the survey on 
the 22 undergraduate students in the first grade of the Department of 
Science Education, Ibaraki University. 
The survey consists of the following two parts:
\begin{itemize}
\item{We conducted a quiz consisting of five questions 
(as for Question 3, the examinees get 1 point only if they answer both (A) and (B) 
correctly) 
from 
contents to be taught in 
junior-high school or first grade of high school. 
This quiz was conducted without announcing in advance. 
Students were to answer them within 15 minutes. 
The detail is given in Appendix \ref{app-test1}.}
\item{We conducted a questionnaire about their history of studying mathematics 
and science, their career preference and their preference of physics, 
whose detail is given in Appendix \ref{app-test2}.}
\end{itemize}
\subsection{Result of questionnaire}
We first report the result of the questionnaire. The detail of the questionnaire 
is given in Appendix \ref{app-test2}.

\paragraph{Q1: History of studying high school subjects} \hspace{0mm} \\
In Question 1, we surveyed the students' history of the following ten high school subjects. 
We give a brief explanation of the Japanese teaching guidelines of mathematics and physics 
in Appendix \ref{app-math}. 
Especially, Math III and Math C are advanced subjects among high school mathematics curriculum.
\begin{center}
 \begin{tabular}{llllll}
 (1) Math I   & (2) Math A     & (3) Math II & (4) Math B        & (5) Math III & (6) Math C \\
 (7) Physics  & (8) Chemistry & (9) Biology & (10) Geoscience   &              &
 \end{tabular}
\end{center}
We asked the following three multiple-choice Yes/No questions for each subject.
\begin{itemize}
\item{(a) Did you study the subject in class?}
\item{(b) Do you think you have acquired this subject?}
\item{(c) Did you study this subject for your entrance 
examinations?\footnote{The entrance examination of Japanese national universities 
(Ibaraki University is one of them) usually consists of two parts: 
(1) National Center Test, which is common to all universities (2) Second-stage exam, 
which is unique to each university. 
The entrance examination of the Department of Science Education, Ibaraki University 
requires the following subjects:
\begin{itemize}
\item{National Center Test (required): Japanese, Geography and History, Civics, 
English, Math I, Math A}
\item{National Center Test (electively required): 
\begin{itemize}
\item{Math: Choose one from (1) Math II (2) Math II, B (3) Industrial Math 
(4) Bookkeeping and Accounting (5) Informatics}
\item{Science: Choose two from 
\begin{enumerate}
\item{One from Integrated Science B and Biology I}
\item{One from Integrated Science A and Chemistry I}
\item{One from Physics I and Geoscience I}
\end{enumerate}}
\end{itemize}}
\item{Second-stage exam: Math I, Math A, Math II, Math B}
\end{itemize}}
}
\end{itemize}
The number of the respondents who answered ``Yes'' is as follows:
 \begin{center}
   \begin{tabular}{|l|l|l|l|l|l|l|l|l|l|l|} \hline
    & 1.MI & 2.MA & 3.MII & 4.MB & 5.MIII & 6.MC & 7.Ph  & 8.Ch & 9.Bio & 10.Geo \\ \hline
(a) & 22 & 22 & 22 & 22 & 12 & 9 & 12 & 21 & 12 & 2 \\ \hline
(b) & 16 & 16 & 11 & 10 &  2 & 1 &  2 &  8 &  4 & 0 \\ \hline
(c) & 21 & 21 & 19 & 19 &  4 & 4 & 12 & 14 &  5 & 2 \\ \hline 
   \end{tabular}
  \end{center}
\begin{itemize}
\item{All of the 22 respondents answered (a)(b), while one of them did not answer (c). 
Thus, the 
number of answers to (c) in the columns of Math I and Math A is not 22 but 21.}
\item{While the Department of Science Education, Ibaraki University does not require 
Math C and Math III, the question (c) refers to the entrance examinations of other universities, 
as well.}
\end{itemize}
This indicates that the students are not confident especially 
in the subjects (5) Math III, (6) Math C and (7) Physics. 
While 12/9/12 students studied (5) Math III, (6) Math C and (7) Physics respectively, 
only 2/1/2 students are confident in these subjects. 
This ratio is especially low compared with other subjects, 
which implies that few students have confidence in these subjects. 

\paragraph{Q2: History of studying undergraduate-level subjects} \hspace{0mm} \\
In Question 2, we surveyed the history of studying the undergraduate-level subjects. Undergraduate-level physics entails advanced undergraduate-level mathematics. We asked multiple-choice Yes/No questions for the following two subjects.
\begin{itemize}
\item{Subjects: (1) Differential and Integral Calculus (2) Linear Algebra}
\item{Items: (a) Did you study the subject in class? (b) Do you think you have acquired this subject?}
\end{itemize}
The number of the respondents who answered ``Yes'' is given as
 \begin{center}
   \begin{tabular}{|l|l|l|} \hline
    & (1) & (2)  \\ \hline
(a) & 9 & 4 \\ \hline
(b) & 1 & 0 \\ \hline
   \end{tabular}
  \end{center}
This indicates that less than half of these students have learned these undergraduate-level mathematics subjects in class. This is partly because the Department of Science Education, Ibaraki University does not require these subjects as compulsory. This result suggests that few of them have confidence in these advanced mathematics subjects.

\paragraph{Q3: Career preference} \hspace{0mm} \\
Their career preference is given below.
 \begin{center}
   \begin{tabular}{|l|l|} \hline  & number \\ \hline \hline
 Elementary school teacher & 7 \\ \hline Junior-high teacher (science) & 7 \\ 
 \hline Junior-high teacher (math) & 2 \\ 
 \hline  High school teacher (science) & 2 \\ \hline High school teacher (math) & 2 \\ 
 \hline Civil servant & 1 \\ \hline Company employee & 1 \\ 
 \hline \hline Total & 22 \\ \hline
   \end{tabular}
  \end{center}
This indicates that many of the students aim for teaching positions, 
especially of science or mathematics.

\paragraph{Q4: Preference of physics} \hspace{0mm} \\
In this question, we surveyed their preference of physics. 
All of the 22 respondents answered this question, and only four of them answered 
that they like physics. 
The period when they have come to like/dislike physics and 
the reason why they like/dislike physics are given below.
{\normalsize
 \begin{center}
   \begin{tabular}{|l|l|l||l|l|l|} \hline Period@& like & dislike & Reason & like & dislike \\ \hline \hline
  Elementary school & & 5 & I can get good scores & 2 & \\ \hline 
  Junior-high school & 1 & & Thanks to cram school & 1 & \\ \hline  
  1st grade of high school & & 2 & Interesting & 1 & \\ \hline 
  2nd grade of high school  & 1 & 3 & Too difficult to understand &  & 13 \\ \hline 
  3nd grade of high school & 2 & 1 &  Mathematically tedious &  & 2  \\ \hline 
  High school (grade not given) & & 2 & I dislike teachers & & 1  \\ \hline 
  University & & 4 & Boring & & 1  \\ \hline
  No answer & & 1 & No answer & & 1  \\ \hline \hline 
  Total & 4 & 18 & Total & 4 & 18 \\ \hline
   \end{tabular}
  \end{center}
}
\noindent
Especially 13 respondents answered that they dislike physics because it is {\it too difficult to understand}. 

\paragraph{Summary of the questionnaire} \hspace{0mm}@\\
In summary, this questionnaire indicates that many of the students 
of the Department of Science Education, Ibaraki University dislike physics. 
The major reason turns out to be that physics is {\it too difficult to understand}. Compared with the other subjects (chemistry or biology), fewer students have confidence in their mastery of physics. We also find that many students have no confidence in advanced mathematics (Math III, Math C or undergraduate-level mathematics), which constitutes a basis of physics.

\subsection{Result of quiz}
In this section, we report the result of the quiz. 
The detail of the questions is given in Appendix \ref{app-test1}. 
These questions are chosen from the 
contents to be taught in 
junior-high school and first grade of high school, 
which are the exam coverage of the teaching certificate 
of natural science in junior-high schools in Ibaraki Prefecture, Japan. 
All of the 22 respondents answered this quiz. 
We categorize the result according to 
the answers to 
the question 1-(7)-(a) of the questionnaire 
in Appendix B.2 (did you study physics in high school?). 
12 of them studied physics in high schools while 10 of them did not. 
The result of the quiz is given below and in figure 1.
 \begin{center}
   \begin{tabular}{|l|l|l|l|} \hline
         & total & studied physics & did not study physics \\ \hline
average@& 2.272 & 2.750 & 1.700 \\ \hline
standard deviation & 1.162 & 1.356 & 0.483 \\ \hline
   \end{tabular}
 \end{center} 
 
  \begin{figure}[htbp]
  \begin{center}
 \includegraphics[scale=0.650]{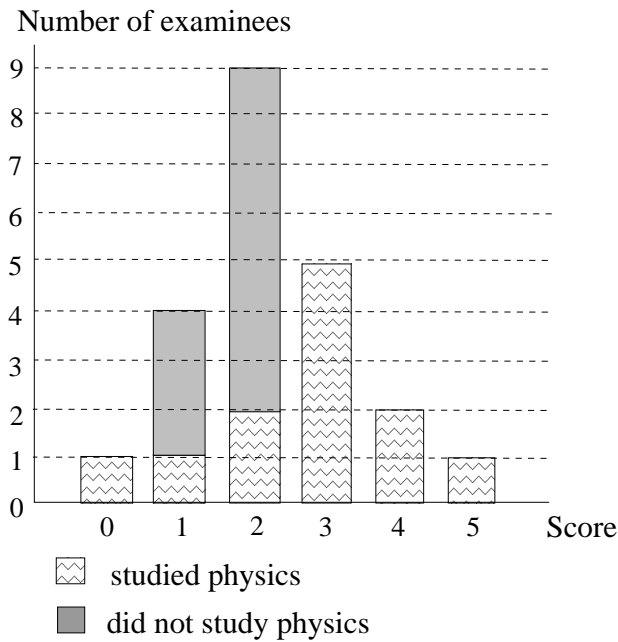}   
  \end{center}
  \caption{The score distribution of the quiz, according to whether 
  they studied physics in high school or not 
  (the item 1-(7)-(a) of the questionnaire in Appendix B.2).}
  \label{test-phys}
  \end{figure}
\noindent
This questionnaire suggests that there is a significant difference 
in their performance of the quiz according to their history of studying physics 
in high schools.

\section{Suggestions to resolve physics unpopularity} \label{chap-plan}
In the previous section, we have seen that most of the undergraduate students 
of the Department of Science Education, Ibaraki University 
have no confidence in physics, 
and very few of them are interested in physics. 
While the quiz and the questionnaire were given to only 22 students, 
we conjecture that this trend 
is seen also in other Japanese university departments of science education. 
Actually, a similar result has been reported in ref.\cite{keisansho} 
about 10,000 Japanese undergraduate students aspiring for science education. 
This leads to the vicious circle of 
the ``reproduction of physics-phobic children by physics-phobic teachers''. 
In this section, we 
make suggestions to improve the present situation.

\subsection{Lecturers by postdoctoral fellows}
Nowadays in Japan, as well as many other countries, 
many Ph.D. holders continue research activities as postdoctoral fellows 
as intermediate steps to obtain academic positions in universities. 
As far as elementary particle theory and nuclear theory are concerned, 
there are 160 Japanese postdoctoral fellows \cite{postdoc}. 
This is a large proportion among the 1,400 Japanese researchers of these two fields. 
A similar phenomenon can be seen in other research fields of natural science.

We arranged a lecture by one of the authors, T.A. 
at College of Education, Ibaraki University 
to enhance the students' interest in physics 
on June 7th 2006
\footnote{At that time T.A. was a postdoctoral fellow at High Energy Accelerator Research 
Organization (KEK), Japan. He will be a permanent lecturer at Setsunan University, Japan in April, 2008.}. 
In Japan, university professors conventionally visit other schools and 
deliver lectures on the cutting-edge science. 
We suggest that postdoctoral fellows also have chances to deliver lectures 
in elementary schools, junior-high schools, high schools or universities. 
Not only the professors but also postdoctoral fellows contribute to 
the frontier of research. 
Actually, postdoctoral fellows have so far produced a lot of exciting results. 
Therefore, it is expected that 
postdoctoral fellows 
can also
deliver lectures 
to convey the enchantment of the cutting-edge science. 
As we mentioned above, there are a large number of postdoctoral fellows. 
Postdoctoral fellows are important human resources 
for such social contributions\footnote{Public lectures for ordinary citizens are 
also important social contributions. 
The authors gave a public lecture on ``the rudiments of elementary particle physics'' 
(which is their research area) at Ibaraki University on June 25th 2006.} 
as delivering lectures outside their institutes.

Our suggestion also gives advantage to postdoctoral fellows themselves. 
Many of them have not so far been given a chance of teaching experience, 
because they already earn their livelihood from the predoctoral or postdoctoral grants. 
In the authors' country Japan, the most typical is the grant of 
JSPS (Japan Science Promoting Society). 
While many of the postdoctoral fellows apply for academic positions in universities, 
these positions are highly competitive especially in natural science. 
In order for them to become strong candidates in such job openings, 
it is often important for them to have experience in teaching. 
The experience of lectures enhances postdoctoral fellows' skill of teaching, 
which will be their important assets in the long run. 

We suggest we deliver lectures even in elementary schools, 
because many of the students have lost interest in natural science 
in their elementary school days. 
Actually, this is reflected in Question 4 of our questionnaire. 
In elementary schools, many teachers themselves 
have little 
interest in natural science. In order to improve this situation, 
we should establish a new system for not only short-term lectures 
but also long-term lectures by ``part-time 
lecturers''.\footnote{At the same time, it is important to establish a system 
to enhance teachers' academic performance of natural science.} 
We should consider a similar system in not only elementary schools 
but also in junior-high schools or high schools. 
Whichever systems are adopted, 
we can resolve the problem of the postdoctoral fellows' shortage of 
educational experience by providing them\footnote{In the authors' country Japan, 
most of the postdoctoral fellows do not have teaching certificate.} 
with workshops of teaching skills.

\subsection{Improvement of mathematical skills}
In studying undergraduate-level natural science, especially physics, 
we need advanced mathematical knowledge. 
However, many of the Japanese university departments of science education only 
require Math I, Math A, Math II and Math B for their entrance 
examinations.\footnote{The entrance examination of Japanese national universities usually consists of 
two parts: (1) National Center Test, which is common to all universities, and 
(2) Second-stage exam, which is unique to each university.} 
The advanced subjects Math III and Math C are not prerequisite.\footnote{Exceptionally, the Department of Science Education of Osaka Kyoiku University 
currently requires Math III and Math C in its Second-stage examination. 
This information was given to T.A. by Masako Tanemura.
The authors would like to thank her for reminding them of this point.} 
Thus, the students are exposed to the undergraduate curricula 
without understanding well Math III or Math C, and have difficulty 
in absorbing undergraduate-level mathematics.

In order to resolve this problem, we would like to make the following three suggestions. 
First is to require Math III and Math C, as well as Math I, Math A, Math II and Math B, 
for the Second-stage exam. 
By doing so, students can establish a solid basis of mathematics and 
enter an undergraduate-level curriculum more smoothly. 
Obviously, the department of science education in general should be categorized 
as the department for the natural science course (rikei), 
rather than 
the humanities or social science course (bunkei), 
since the students in the department need to study advanced science 
in order to become science teachers in the future. 
Therefore, the advanced mathematics, Math III and Math C, 
should be universally required for the entrance examinations of
university departments of science education. 

Second is to make undergraduate-level mathematics compulsory. 
As far as the Department of Science Education, Ibaraki University, is concerned, 
such advanced mathematics classes are not compulsory. 
At least, we should require Differential and Integral Calculus and Linear Algebra 
as compulsory subjects. 
We believe this arrangement enhances students' mathematical ability and hence helps students understand advanced science.

\begin{figure}[htbp]
  \begin{center}
 \includegraphics[scale=1.00]{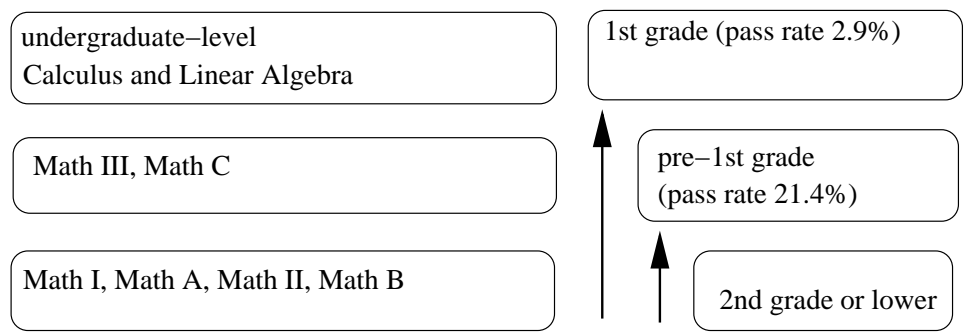}
  \end{center}
  \caption{The exam coverage and the pass rates of 1st and pre-1st grades of Suken 
  (Practical Mathematics Proficiency Test) for the 2004 academic year.}
  \label{suken-fig}
  \end{figure}

Third is to recommend the students to take 
``Suken (Practical Mathematics Proficiency Test)''\cite{suken-ref}, 
which is administered by ``The Mathematics Certification Institute of Japan''. 
%
%
This examination had been held only in the authors' country Japan originally, 
but now it is conducted also in the United States, 
South Korea, Singapore, Indonesia, etc. 
We guess that there may be similar examinations in other countries.
The exams consist of ten grades; 1st, pre-1st, 2nd, pre-2nd, 
and 3rd 
to 
8th grades. 
Especially, we recommend the students aspiring for teaching science 
to take the 1st (pre-1st) grades, whose exam coverage is 
at most undergraduate-level Calculus and Linear Algebra (Math III and Math C), 
respectively. 
The exam coverage and the pass rates are given in figure \ref{suken-fig}.
  
We believe ``Suken'' enhances the students' motivation to study mathematics. 
Students can recognize their achievement in their scores and the success 
in certain grades. 
``Suken'' is the only mathematical certification 
in Japan to 
assess undergraduate-level mathematics. 
Thus, ``Suken'' is an important catalyst to motivate the students 
to study advanced mathematics. 
In Japan, many universities give credits to students who have reached a certain level of Eiken 
(Test in Practical English Proficiency)\footnote{This exam is administered by 
the Japanese organization STEP (Society for Testing English Proficiency). 
The official cite is {\tt http://stepeiken.org/} .}, TOEIC or TOEFL. 
On the other hand, giving credits for passing Suken is not 
so prevalent.\footnote{The list of the schools giving credits for passing Suken 
is given in {\tt http://www.suken.net/treat/nintei200606.html} .} 
It is worth while to consider giving credits for passing Suken or requiring Suken 
in the entrance examination of 
graduate course\footnote{Currently, the Department of Science Education, 
Ibaraki University does not require mathematics in its entrance examination 
of graduate course.} in the university departments of science education.

\subsection{How to cure teachers' ``physics (science)-phobia''}
Next, we would like to make suggestions to resolve the unpopularity of natural science, 
especially physics, among the teachers. 
First is to increase the ratio of natural science (physics) 
in the examination of teaching certificate. 
We also need to consider raising the passing score. 
Second is to increase 
the compulsory subjects in the undergraduate course in science education. 
Third is to introduce the renewal of 
the teaching certificate to assess the current teachers' scholastic performance 
and the education program of science and physics.

Currently in Japan, the SPP (Science Partnership Program), which is administered by 
JST (Japan Science and Technology Agency) \cite{JST}, 
contributes to enhancing the teachers' scholastic ability of natural science, 
where university professors provide the teachers of elementary and secondary education 
with workshops. 
While such workshops contribute to raising the teachers' interest 
in natural science temporarily, they cannot resolve the deep-rooted problem in the long run. 
It is impossible to raise their scholastic ability only 
with two- or three-hour workshops. 
If it were possible, science-phobic children or students could improve their scholastic 
ability of science without any difficulties. 
On the other hand, some teachers obtain specialized teaching certificate 
via a graduate course in science education. 
This route, however, does not contribute 
to resolving the problem of teachers' academic performance, either. 
First, with a recommendation of Education Committee, 
they are practically admitted into the graduate school without entrance examinations.
Second, they do not have a solid assessment of their academic ability of 
natural science while graduate students.

Then, what can we do to improve the teachers' ability of natural science? 
The only solution is that teachers themselves spend enough time 
reviewing natural science of junior-high school, high school or undergraduate course. 
There is no royal road to learning. 
It is much more effective to solve exercises of the textbooks step-by-step than 
to attend classes of the SPP. 
In short, the most effective solution to enhancing the teachers' scholastic ability is 
to ``introduce a strict renewal system of teaching certificate and 
require the exams of natural science'', in which teachers cannot renew their license 
without passing the test. Of course, the system must allow the 
disqualified teachers to retry the renewal exams. 
Also, 
some programs 
to encourage the teachers' self-studying should be developed.

We believe the only solution is to ``require exams of natural science 
in a rigid renewal system of teaching certificate''. 
Actually, in Japan it is highly controversial 
whether we should introduce a renewal system of teaching certificate, 
which may be an important step toward a rigid assessment. 
However, without introducing a rigid assessment, the renewal system would not 
work for resolving the teachers' science- and physics-phobia 
and improving their educational ability.
We are urged to introduce such systems, which would be beneficial 
for the citizens in the long run.

\section{Conclusion} \label{chap-conclu}
In this paper, we have discussed how to resolve 
``the reproduction of physics-phobic children by physics-phobic teachers'', 
which is one of the reasons for the unpopularity of natural science among children. 
To that end, we have conducted a quiz and questionnaire 
on the 22 undergraduate students in the first grade 
of the Department of Science Education, Ibaraki University. 
We have found that many of these students dislike physics and 
have difficulty in understanding physics. 
We have made the following three suggestions to improve such a situation. 
First is to encourage the lectures of postdoctoral fellows in elementary or 
secondary education. 
We arranged a lecture by one of the authors, T.A. 
at the undergraduate course of the Department of Science Education, Ibaraki University, 
when he was a postdoctoral fellow at High Energy Accelerator Research Organization (KEK), 
Japan. 
We have also proposed a system in which postdoctoral fellows deliver 
lectures as part-time teachers in elementary schools. 
We need to do likewise in junior-high schools or high schools. 
To make use of postdoctoral fellows for science education is beneficial 
not only for science education but also for the postdocs themselves. 
Second, we have suggested the enhancement of mathematical performance. 
To that end, we have recommended Suken (Practical Mathematics Proficiency Test), 
to motivate the students to study mathematics by themselves. 
Third, we have made three suggestions to resolve 
the unpopularity of natural science among the teachers. 
Two of them concern the teaching certificate examination and 
the curricula of the university departments of science education. 
The other is to ``introduce a rigid renewal system of teaching certificate'' 
for the current teachers. 
We believe this is the only solution to enhancing the teachers' scholastic ability 
of natural science.

We now live in a society based on advanced scientific technology, 
and we are involved in scientific technology throughout our lives. 
In such a society, natural science, 
which constitutes the basis of scientific technology, is prerequisite knowledge.
If the unpopularity of natural science accelerates in the future, 
it may affect the fate of the whole nation. In order to inherit the culture of 
natural science to the next generation, we must resolve the problem of 
the unpopularity of natural science.

 \paragraph{Acknowledgment} \hspace{0mm} \\
The authors would like to thank the students of College of Education, 
Ibaraki University, who cooperated in the questionnaire and attended the lecture. 
They also hope that they become more and more interested in physics. 
They would like to thank Spenta R. Wadia for carefully reading the manuscript. 

\appendix
\section{Japanese teaching guidelines of mathematics and physics} \label{app-math}
In this appendix, we present the detail of Japanese teaching guidelines in high school. 
These guidelines are summarized in the website 
of MEXT (Ministry of Education, Culture, Sports, Science and Technology, Japan) 
\cite{shidou}, which is written in Japanese. 
In Japan, high school mathematics is divided into 
the following six subjects: Math I, Math A, Math II, Math B, Math III and Math C. 
Especially, Math III and Math C are advanced subjects, and usually in Japan 
the high school students advancing to 
the university departments of ``the humanities or social science course (bunkei)'' 
do not learn Math III and Math C. 
College of Education, which the department of science education belongs to, 
is usually categorized as ``bunkei'' in Japan.
\begin{itemize}
\item{Math I: expansion and factorization, linear and quadratic equation, quadratic function, trigonometry}
\item{Math A: geometry of plane figures, set and logic, probability and binomial coefficients}
\item{Math II: proof of equalities and inequalities, complex plane, cubic or higher-degree equation, coordinate and plane figures, differential and integral calculus of polynomials}
\item{Math B: numerical sequence, vector, statistics and computer}
\item{Math III: differential and integral calculus of miscellaneous functions (trigonometric, logarithmic and exponential functions)}
\item{Math C: matrix, quadratic curves, probability distribution}
\end{itemize}

The curriculum of natural science in Japanese high school is divided into 
the following eleven subjects: Rudimentary Science, Integrated Science A, 
Integrated Science B, Physics I, Physics II, Chemistry I, Chemistry II, 
Biology I, Biology II, Geoscience I and Geoscience II. 
Especially, physics is taught in Rudimentary Science, Integrated Science A, 
Physics I and Physics II. The detail of the physics curriculum is as follows:
\begin{itemize}
\item{Rudimentary Science: evolution of science, composition of materials, origin of life, energy, earth and space}
\item{Integrated Science A: nature and energy, material and human life, development of science and technology}
\item{Physics I: electricity, acoustic and optical waves, potential and kinetic energy}
\item{Physics II: momentum and energy, harmonic oscillation, universal gravity, electric and magnetic fields, motion of atoms and molecules, composition of atoms}
\end{itemize}

\section{Detail of the quiz and questionnaire} \label{app-test}
\subsection{Questions of the quiz} \label{app-test1}
Answer the following questions. (5 points, in total) \\

\noindent
{\bf Question 1} We fill a syringe with ideal gas of pressure 
$1.0 \times 10^{5}$ [Pa] and volume 50 [cm$^3$], 
and close the pointy end of the syringe. 
We pull up the piston of the syringe and expand this ideal gas to 100 [cm$^3$]. 
In this process the temperature of the gas remains constant. Calculate the pressure of this ideal gas. (1 point) \\

\noindent
{\bf Question 2} We shoot the bullet with the mass $m$ [kg] and initial velocity $v$ [m/s] toward the block of mass $M$ [kg] hung by the thread. The bullet remains in the block after the collision. Answer the following questions. Here, the gravitational constant is $g$ [m/s$^2$]. (1 point for each, 2 points in total)
\begin{enumerate}
\item{Suppose the bullet and the block move at the velocity $v'$ [m/s] just after the bullet collides with the block. Express $v'$ using $m,M,v$. }
\item{The bullet and the block rise to the maximum height of $h$ [m] together. Express $v$ using $m,M,g,h$.}
\end{enumerate}
  \begin{figure}[htbp]
   \begin{center}
    \scalebox{0.70}{\includegraphics{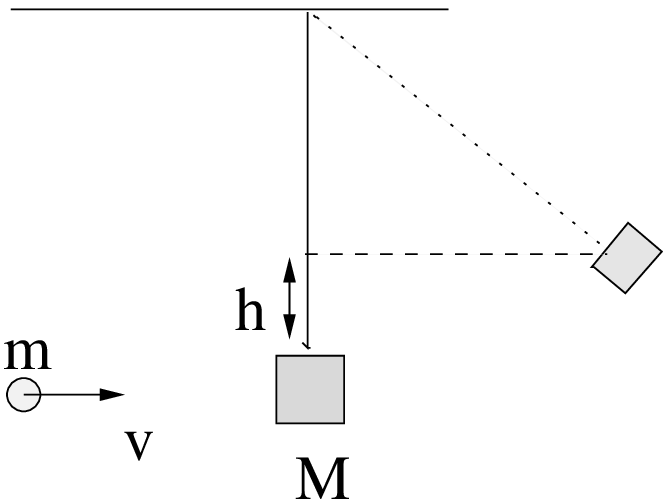} }
    \end{center}
  \end{figure}

\noindent
{\bf Question 3} Fill the blanks (A) and (B) with the most appropriate words. (1 point only for both (A) and (B)) 
\begin{itemize}
\item{We hang a buzzer with a thread inside a bottle. We remove the air in the bottle using a vacuum pump. The sound of the buzzer (A). \\
1.remains constant \hspace{1mm} 2.becomes larger gradually \hspace{1mm} 3.becomes smaller gradually}
\item{When we beat a drum near the flame of a candle, the flame (B). \\
1.waves \hspace{1mm} 2.does not wave}
\end{itemize}

\noindent
{\bf Question 4} Calculate the electric current in the 3.0 $\Omega$ resistor in the circuit shown below. (1 point)
  \begin{figure}[htbp]
   \begin{center}
    \scalebox{0.70}{\includegraphics{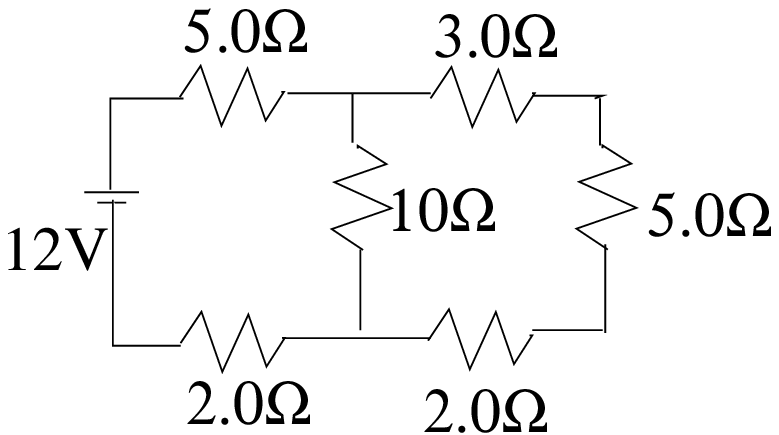} }
    \end{center}
  \end{figure}

\vspace*{-1.0cm}

\noindent
{\bf Answer} \\ 
Q1: 5.0 $\times$ 10$^4$[Pa], \hspace{1mm} Q2: 1.$v'=\frac{mv}{m+M}$, 2.$v= \frac{M+m}{m} \sqrt{2gh}$, \hspace{1mm} Q3: (A)3(B)1, \hspace{1mm} Q4: 0.50[A].

\subsection{Questionnaire} \label{app-test2}

\begin{enumerate}
\item{Answer the following questions about the high school subjects. \\
(1) Math I \hspace{1mm} (2) Math A \hspace{1mm}  (3) Math II \hspace{1mm}  
(4) Math B \hspace{1mm}  (5) Math III \hspace{1mm}  (6) Math C \hspace{1mm}  \\ 
(7) Physics \hspace{1mm}  (8) Chemistry \hspace{1mm} (9) Biology \hspace{1mm} (10) Geoscience
\begin{enumerate}
\item{Did you study the subject in class? [Y/N]}
\item{Do you think you have acquired this subject? [Y/N]}
\item{Did you study this subject for your entrance examinations? [Y/N]}
\end{enumerate}
}
\item{Answer the following questions about the undergraduate-level subjects. \\
(1) Differential and Integral Calculus \hspace{1mm} (2) Linear Algebra
\begin{enumerate}
\item{Did you study the subject in class? [Y/N]}
\item{Do you think you have acquired this subject? [Y/N]}
\end{enumerate}
}
\item{Answer your career preference. \\
1. Teacher \hspace{1mm} 2. Civil servant \hspace{1mm} 3. Company employee \hspace{1mm} 
4. Other
\begin{itemize}
\item{If you answer 1., answer which you opt for, ``elementary school'', ``junior-high school'', ``high school'' or ``university''.}
\item{If you choose ``elementary school'', answer which you opt for [general / music or housecraft].}
\item{if you choose ``junior-high school'', ``high school'' or ``university'', 
answer the subject you want to teach.}
\end{itemize}
}
\item{Answer the following questions about your preference of physics.
\begin{enumerate}
\item{Do you like physics? [Y/N]}
\item{When have you come to like/dislike physics?}
\item{Why do you like/dislike physics?}
\end{enumerate}
}
\end{enumerate}

\end{document}